\documentclass[%
 amsmath,amssymb,
 aps,
pra,
dvipdfmx,
twocolumn,
]{revtex4-1}

\usepackage{algorithm}
\usepackage{algpseudocode}
\usepackage{graphicx}
\usepackage{dcolumn}
\usepackage{bm}

\begin{document}

\preprint{APS/123-QED}

\title{A Fidelity Susceptibility Approach to Quantum Annealing of NP-hard problems}

\author{Jun Takahashi}
 \email{jt@huku.c.u-tokyo.ac.jp}
\author{Koji Hukushima}%
 \email{hukusima@phys.c.u-tokyo.ac.jp}
\affiliation{%
Department of Basic Science, University of Tokyo\\
3-8-1 Komaba, Meguro, Tokyo 153-8902, Japan
}%

\date{\today}

\begin{abstract}
The computational complexity conjecture of 
$\mathrm{NP} \nsubseteq \mathrm{BQP}$ 
implies 
that there should be an exponentially small energy gap
for Quantum Annealing (QA) of NP-hard problems.
We aim to verify how this computation originated gapless point could be understood based on physics, using the quantum Monte Carlo method.
As a result, we found a phase transition detectable only by the divergence of fidelity susceptibility.
 The exponentially small gapless points of each instance are all located
 in the phase found in this study, 
which suggests that this phase transition is the physical cause of the failure of QA for NP-hard problems.
\begin{description}
\item[PACS numbers]
03.67.Ac, 03.67.Lx, 64.70.Tg, 75.10.Nr
\end{description}
\end{abstract}

\pacs{Valid PACS appear here}
\maketitle


\section{Introduction}
Recent progress in quantum technology 
has greatly accelerated interests towards quantum computation. 
One of the promising techniques to use quantum effects for computation is 
Quantum Annealing (QA) \cite{Kadowaki-Nishimori,Farhi,DWAVE}.
QA uses quantum fluctuation
induced by a Hamiltonian $\hat{H}_{\mathrm{T}}$,
which is usually a simple transverse field $\sum_i \hat{\sigma}_i^x$.
By constructing a problem Hamiltonian $\hat{H}_{\mathrm{P}}$,
whose 
ground state encodes the answer of
the problem in interest,
the total Hamiltonian is
\begin{equation}
\hat{H}(\lambda)=(1-\lambda)\hat{H}_{\mathrm{P}}+\lambda\hat{H}_{\mathrm{T}},
\end{equation}
where the parameter $\lambda$ is the time dependent control parameter.
The sufficient condition of how slow 
the quantum fluctuation should be reduced
is closely related to the minimum excitation gap $\Delta E_{\min}$ of $\hat{H}(\lambda)$.
The quantum adiabatic theorem \cite{quantumadiabaticthm}  
guarantees 
that the quantum state starting 
from the ground state of $\hat{H}_{\mathrm{T}}$ will always remain 
in the instantaneous ground state with high probability,
as long as $\lambda$ is changed using time longer than ${O}(\Delta E_{\min}^{-2})$.
Thus, for a fixed problem embedded in 
$\hat{H}_{\mathrm{P}}$,
if there are only polynomially small energy gaps, 
it indicates that QA can solve the specific problem in polynomial time, 
which is considered as ``efficient'' in the computer science community \citep{arora-barak,natcom}. 
Early studies hinted polynomial time computation of NP-hard problems \cite{SolveNPc}, 
however they turned out to be artifacts of  finite-size effects 
\cite{SpinGlassIsReason}.
After numerical supports that
naive QA fails for NP-hard problems were found,
some scenarios such as Anderson/many-body localization or spin glass transition
\cite{AvoidedCrossingIsReason,AndersonLocalizationIsReason,SpinGlassIsReason}
were proposed to explain the reason of this failure. 
The wide belief in the computer science community
that even quantum computers would not be able to solve 
NP-hard problems in polynomial time 
(NP$\nsubseteq$BQP)\cite{npnsubsetbqporacle},
could be seen as a {\it computational complexity based} physical conjecture 
that says all $\hat{H}(\lambda)$ should have an exponentially small gapless point at some $\lambda$
if $\hat{H}_{\mathrm{P}}$ corresponds to a NP-hard problem.
Providing a physical picture to this conjecture would be of great importance 
for the understanding of the connection between 
computational complexity and physics.
One of the convincing arguments was made using numerical approaches \cite{APYoungFOT}, 
showing that a fraction of samples exhibit first order phase transition-like behaviors
in terms of the usual spin-glass order parameter $q$. 
Since first order transitions in quantum spin systems are usually accompanied
with an exponentially small energy gap \cite{FOTisEXPmeanfield}, 
they provide good reason 
that 
NP-hard problems could not be computed in polynomial time using QA.
However, those first order transitions are strongly sample dependent, 
and no concrete arguments have been made 
for connecting the phenomena to the above mentioned scenarios.

In this work, we see if those ``underlying phase transition'' scenarios are correct at all,
using quantum Monte-Carlo simulations.
A natural strategy to see the underlying phase transition 
would be to take the sample average of the 
the spin-glass order parameter $q$ 
which exhibited a first order transition-like behavior in individual samples.
However, we find that this quantity
does not exhibit any singularities when the sample average $\bar{q}$ is taken. 
This implies that another measure is necessary 
to see the phase transition 
in question,
 if it exists.
We thus used the notion of 
{\it fidelity susceptibility} \cite{chiF1,Fidelity-gu,Fidelity-zanardi}.
This quantity quantifies how rapid the ground state is changing in the $\lambda$ direction.
The fidelity susceptibility $\chi_{\mathrm{F}}$ is also proportional to
the symmetric logarithmic derivative (SLD) Fisher information metric\cite{SLDFisher},
and has been recently under interest 
for detecting phase transitions where the order parameter is unknown,
such as topological order phases.
Our work suggests that $\chi_{\mathrm{F}}$ is also useful for quantum spin-glass like models with quenched disorders. 

By using stochastic series expansion (SSE), 
a variant of quantum Monte Carlo methods, 
we estimate the sample average of the 
fidelity susceptibility $\overline{\chi_{\mathrm{F}}}$. 
We show for a specific NP-hard problem that  
QA undergoes a phase transition at a certain value of $\lambda$.  
At this transition point, $\overline{\chi_{\mathrm{F}}}$ diverges,
while other common
quantities such as 
$\bar{q}$ does not show or has very weak singularity.
This implies that although there is a quantum phase transition at the
value of $\lambda$, 
the order parameter for this transition is yet unknown.
However, since all of the first order like transitions occur at values below
the transition point, 
this result suggests that the first order transitions could be understood
as a phenomenon within a non-trivial quantum phase.
The paper is organized as follows. 
First we explain our model, which is a specific NP-hard problem that we fix.
Then we explain the numerical methods used in this work, together with how to 
measure the fidelity susceptibility $\chi_{\mathrm{F}}$ for this system.
Section \ref{sec:results} explains our main results, and finally we will conclude our work.

\section{MODEL: MAXIMUM INDEPENDENT SET}
We fix the NP-hard problem in consideration 
to the Maximum Independent Set (MIS) problem. 
This is the problem where given a graph $G=(V,E)$, 
one finds the maximum subset of vertices $I^*\subset V$ 
such that no two vertices in $I^*$ are adjacent 
(i.e. $\forall i, j \in I^*, (i,j)\notin E$).
Finding the solution of the MIS problem is equivalent to 
finding the ground state of a Hamiltonian with the Pauli matrix in the form of
\begin{equation}
\hat{H}_{\mathrm{P}}=\frac{c}{4}\sum_{(i,j)\in E}\hat{\sigma}_i^z\hat{\sigma}_j^z
-\sum_i\frac{2-cd_i}{4}\hat{\sigma}_i^z
\end{equation}
where $d_i$ is the degree of vertex $i$ in graph $G$
and we set $c>1$ to be 2.
We will generate random instances of the MIS problem
by randomly generating the graph $G$ in target.
We first generate an 
Erd{\"o}es-R{\'e}nyi random graph \cite{ERgraph} 
and then randomly add extra edges 
to it 
in order to make the ground state unique. 
The details are explained 
in Appendix A.
No degenerate ground state yields that 
the first excited gap $\Delta E$
surely determines the computation time.
Another reason is that the first order transitions mentioned in the introduction are 
previously only observed in those examples 
which have unique solutions\cite{APYoungFOT}.
We should care that 
making the solution unique
does not affect the {\it hardness} of the problem,
and we describe 
the details of how this point is taken care of in Appendix B.
We also confirm from \cite{zhouVCRSB} 
that the classical MIS of Erd{\"o}s-R{\'e}nyi random graphs with average degree $d>e=2.718...$
is in the replica symmetry breaking (RSB) phase,
and therefore use Erd{\"o}s-R{\'e}nyi random graphs with average degree $d=3$ to start with.
Adding edges to make the solution unique 
only effectively increases the average degree,
and pushes the graph deeper into the RSB phase.

\section{METHOD: SSE WITH REPLICA EXCHANGE AND FIDELITY SUSCEPTIBILITY}
We adopt the SSE method for Ising spin systems \cite{Sandvick1,sandvickloop}.
The SSE method effectively takes the Trotter limit in the 
path integral Monte Carlo method\cite{Suzuki,Trotter}, 
and is therefore free from systematic error caused by the Trotter decomposition.
Assuming that the Hamiltonian $\hat{H}$ is decoupled into some local
operators as $\hat{H}=-\sum_k \hat{W}_k$  and that 
with an appropriate basis $B=\{|\sigma_m\rangle\}_m$, the non-negative hopping 
elements $W_{k,m}$ for $\forall k, m$ are expressed as 
$\hat{W}_k|\sigma_m\rangle =W_{k,m}|\sigma_{m^{\prime}}\rangle$ 
with 
$|\sigma_{m^{\prime}}\rangle \in B$, 
the partition function $Z$ is expanded as 
\begin{eqnarray}
Z&:=&\mathrm{Tr}[e^{-\beta\hat{H}}]\\
&=&\sum_{n=0}^{\infty}\frac{\beta^n}{n!}\sum_{\{k_l\}}\sum_m 
\langle \sigma_m| \prod_{l=1}^{n} \hat{W}_{k_l} |\sigma_m\rangle .
\end{eqnarray}
SSE samples 
terms  in the above summation using Markov chain Monte Carlo methods,
by changing $|\sigma_m\rangle$, $n$ and $\{\hat{W}_{k_l}\}_l$.
For the present model, we 
take the $z$-basis for $B$ 
and simply have every $(1-\lambda)J_{ij}\hat{\sigma}_i^z\hat{\sigma}_j^z$ terms, 
every $(1-\lambda)h_i\hat{\sigma}_i^z$ terms, 
and every $\lambda\hat{\sigma}_i^x$ terms as $\hat{W}_k$.
By adding constants to each term, one could make all $W_{k,m}\geq 0$.
We adopt the usual Swendsen--Wang type procedure for the global update, 
but with a minor modification due to the presence of local fields.
We adjust the constant terms associated with the $(1-\lambda)h_i\hat{\sigma}_i^z$ terms,
and make the corresponding $W_{k,m}=0$ or $2(1-\lambda)h_i$. 
This allows us to treat those operators as same as other ones
except for the global updates, 
where we never flip clusters which contain those operators.

Furthermore, we also adopt the exchange Monte Carlo method (EMC) 
to accelerate equilibration \cite{hukushima-nemoto, itayhen}.
We divide the parameter region $[\lambda_{\mathrm{low}},\lambda_{\mathrm{high}}]$ into
$R-1$ equidistributed intervals,
and run $R$ different SSE simulations with corresponding $\lambda_r$.
Configurations of adjacent $\lambda_r$ are exchanged with 
probability
\begin{equation}\nonumber
\min\left[~1, ~
\left(\frac{\lambda_r}{\lambda_{r+1}}\right) ^{T_{r+1}-T_r}
\left(\frac{1-\lambda_r}{1-\lambda_{r+1}}\right) ^{P_{r+1}-P_r}
\right]~,
\end{equation}
where $T_{r}$ and $P_r$ denote the number of operators coming from 
$\hat{H}_{\mathrm{T}}$ and $\hat{H}_{\mathrm{P}}$, respectively.
This ensures that the over-all $R$ runs will 
satisfy the detailed balance condition,
resulting in acceleration of equilibration.

Since we take the strategy of observing the ground-state properties by
sampling the equilibrium state with high enough $\beta$,
we should extend the definition of fidelity to finite-temperature states 
as in \cite{prxmadesimple},
\begin{equation}
F(\lambda,\lambda+\epsilon)=
\sqrt{\frac{\mathrm{Tr}[e^{-\beta\hat{H}(\lambda)/2}e^{-\beta\hat{H}(\lambda+\epsilon)/2}]}
{(\mathrm{Tr}[e^{-\beta\hat{H}(\lambda)}]\mathrm{Tr}[e^{-\beta\hat{H}(\lambda+\epsilon)}])^{1/2}}}.
\label{eq:chiF}
\end{equation}
By expanding Eq.~(\ref{eq:chiF}) up to the second term in $\epsilon$, 
we obtain the representation of $\chi_{\mathrm{F}}$ for finite temperature as
\begin{eqnarray}
\chi_{\mathrm{F}}^{T\neq 0}=
\frac{\langle k_L k_R\rangle -\langle k_L \rangle \langle k_R\rangle }{2\lambda^2}
+\frac{\langle l_L l_R\rangle -\langle l_L \rangle \langle l_R\rangle }{2(1-\lambda)^2}
\nonumber\\
-\frac{\langle k_L l_R\rangle -\langle k_L \rangle \langle l_R\rangle }{2\lambda(1-\lambda)}
-\frac{\langle l_L k_R\rangle -\langle l_L \rangle \langle k_R\rangle }{2\lambda(1-\lambda)},
\end{eqnarray}
where $k$ and $l$ denote the number of operators
within $\{\hat{W}_{k_l}\}_{l=1}^{n}$
coming from $\hat{H}_{\mathrm{T}}$, 
and $\hat{H}_{\mathrm{P}}$, respectively, and 
$\langle \cdots \rangle$ represents the Monte Carlo average.
The subindices $L$ and $R$ represent the number of the according operators
in the left half or the right half of the operator string $\{\hat{W}_{k_l}\}_{l=1}^{n}$ .
The center of the string is determined probabilistically for 
every
sampled configuration, 
according to the binomial distribution
among the possible $n+1$ points of division.

\section{NUMERICAL RESULTS}
\label{sec:results}
By using the SSE method together with EMC, 
we sample the equilibrium state of $\beta = 3.5N$ with $N$ being the
number of vertices of the problem. 
It is known that whenever $\beta > \Delta E_{\min} ^{-1}$, 
the equilibrium state is close to the ground state \cite{APYoungFOT,Yasuda-Suwa-Todo}.
By fixing the system size $N$ and increasing $\beta$,
we see that usual observables saturate around $\beta \sim 1.5N$
and $\overline{\chi_{\mathrm{F}}}$ also saturates at $\beta\sim 3.5N$.
Thus, sampling the thermal equilibrium state at $\beta = 3.5 N$
is 
sufficient 
to see the properties of the ground state.
3072 samples were taken for system sizes $N=20$ and 30, 
and $1.6\times 10^5$ to $1.28 \times 10^6$ MCS to measure the quantities after the same amount for equilibration.
In the following, we show 256 samples with $N=50$ since they exhibit 
clearer first order transitions, but will not use them for sample averages due to lack of enough samples. 

\subsection{Spin-glass order parameter $q$}
\begin{figure}[t]
\includegraphics[width=9cm]{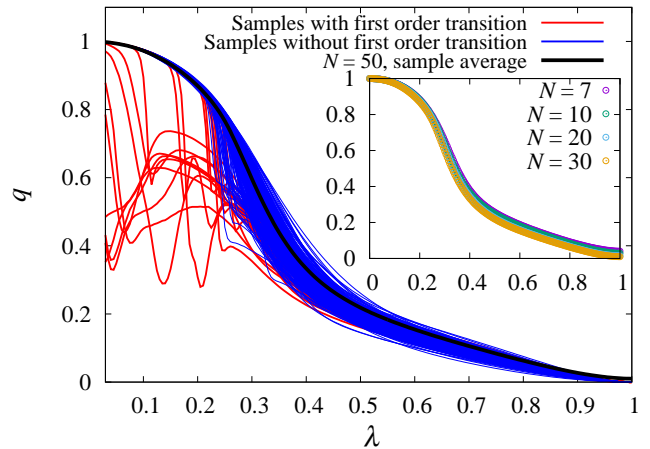}
\caption{\label{fig:q} The overlap parameter $q$ is drawn as a function
 of $\lambda$ for 256 samples with $N=50$. 
Samples with dips are drawn in red, and samples without dips are in blue.
The black line is the sample average $\bar{q}$. 
 The inset presents $\bar{q}$ for different sizes. 
 }
\end{figure}

The spin-glass order parameter $q$ (the overlap parameter) is defined as 
\begin{equation} 
q:=\frac{1}{N}\sum_i \langle \hat{\sigma}_i^{(1)} \hat{\sigma}_i^{(2)}\rangle 
=\frac{1}{N}\sum_i \langle \hat{\sigma}_i \rangle ^2,
\end{equation}
where the upper suffixes ${(1)}$ and ${(2)}$ are the labels for two independent systems with the same quenched disorder.
In previous study \cite{APYoungFOT}, 
some samples of NP-hard problems exhibited an acute increase in $q$
after a characteristic dip with decreasing $\lambda$ 
and this phenomena was called first order phase transitions after a physical argument.
First order phase transitions are notions which are well-defined only in the thermodynamic limit,
however we will abuse the term in this paper following Ref.~\cite{APYoungFOT}.
These first order transitions are also found in our model as in Fig.~\ref{fig:q},
and we confirmed that double peaks in the histogram of $q$ and
$F_{\mathrm{ans}}$ defined below 
 were observed in the ``transition points" of those ``first order transitions",
supporting the argument of Ref.~\cite{APYoungFOT}.
\begin{figure}[h]
\includegraphics[width=9cm]{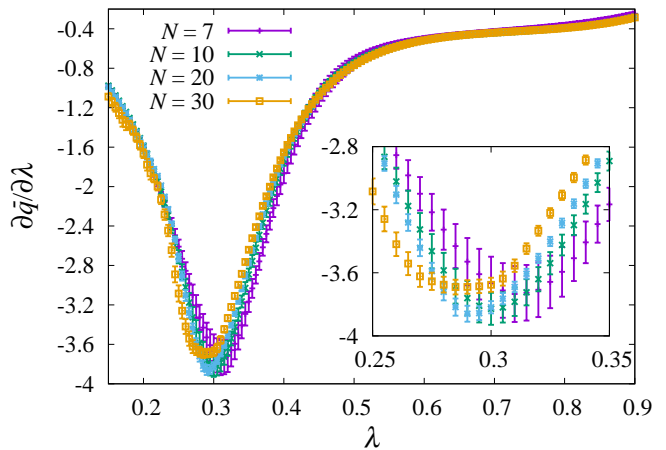}
\caption{\label{fig:dqdl}The $\lambda$ derivative of the averaged
 spin-glass order parameter $\bar{q}$ for different sizes are drawn for
 different sizes. 
The inset is a magnified view at the 
minimum point.
}
\end{figure}

The transition points and even the presence of the first order transitions 
are sample dependent.
When we take the sample average of $q$ to see the behavior of the ensemble
as in Fig.~\ref{fig:q} (inset), 
we are unable to see any size dependence nor singularities.
The derivative with respect to $\lambda$, 
$\partial q/ \partial \lambda$,  also seems to have no singularities
as seen in Fig.~\ref{fig:dqdl}.
This implies that although $q$ was a good quantity for detecting first
order phase transitions for individual samples
in QA, 
it does not capture the underlying phase transition for this problem.

\subsection{Fidelity susceptibility $\chi_{\mathrm{F}}$}
\begin{figure}[h]
 \includegraphics[width=9cm]{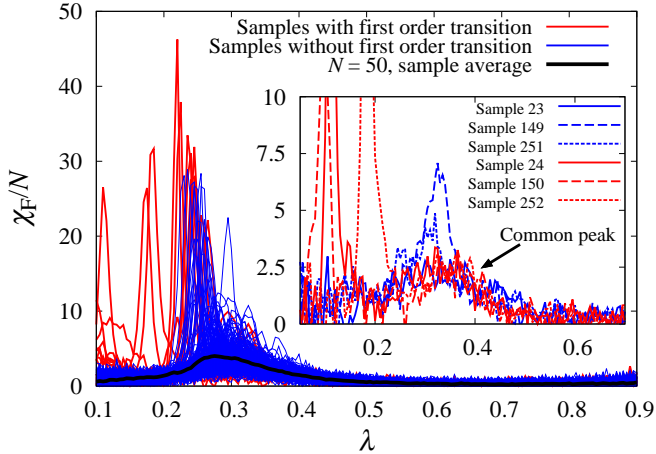}
 \caption{\label{fig:chiF_takusan}
 $\lambda$ dependence of the fidelity susceptibility $\chi_{\mathrm{F}}$
 for 256 samples with $N=50$. 
The use of colors is the same as Fig.~\ref{fig:q}.
The inset shows a magnified view of 6 samples for clarity.
The arrow is pointing at the moderate peak of $\chi_{\mathrm{F}}$ at $\lambda\sim 0.3$}
\end{figure}
\begin{figure}[h]
\includegraphics[width=9cm]{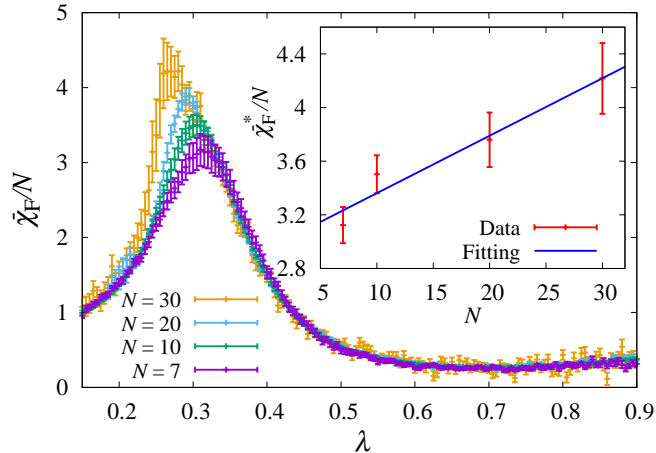}
\caption{\label{fig:chiF}The sample average of normalized fidelity susceptibility $\overline{\chi_{\mathrm{F}}}/N$ is drawn for different system sizes. 
The inset shows peak values and a least square fitting.}
\end{figure}

Fig.~\ref{fig:chiF_takusan} presents $\lambda$ dependence of the
fidelity susceptibility $\chi_F$ for each sample with $N=50$.
Similarly to $q$, 
 $\chi_{\mathrm{F}}$ also shows an acute peak
at the points where first order transitions occur (See Fig.~\ref{fig:chiF_takusan}).
Other than the acute peaks corresponding to the first order transitions,
we can see a relatively moderate peak at $\lambda \sim 0.3$.
Importantly, this moderate peak is present in samples of both types,
either with or without first order transitions, as seen in the inset of Fig.~\ref{fig:chiF_takusan}.
This indicates that there may be a phase transition
which is different from the first order transitions previously discussed.

To see this more clearly, 
the sample average of the fidelity susceptibility
$\overline{\chi_{\mathrm{F}}}$ is taken in Fig.~\ref{fig:chiF}. 
It shows a diverging trend towards the thermodynamic limit.
It should be noted that all of the first order transitions occur within the low $\lambda$ phase of this transition,
which indicates that this phase causes the first order transitions.

We do not understand yet what type of phase transitions this is,
since they do not accompany any singularities in either $q$, nor $S_{\mathrm{ans}}$,
a R{\'e}nyi entropy-like quantity 
which will be explained in the following section.

\subsection{Answer Fidelity}
We can easily measure the fidelity between 
the ground state of a particular $\lambda$ and that of $\lambda=0$,
since we make the solution of MIS unique.
We will call this as the answer fidelity $F_{\mathrm{ans}} := F(\lambda,0)$,
and this will represent the (square root of) the probability of
observing the correct answer by a projection measurement 
with the $z$-basis, $\langle \hat{P}_{\mathrm{ans}}\rangle$.

\begin{figure}[t]
\includegraphics[width=9cm]{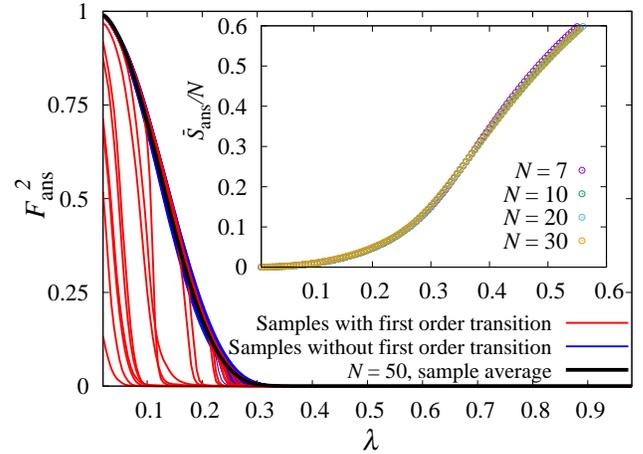}
\caption{\label{fig:Fans} Square of the answer fidelity $F_{\mathrm{ans}}^2=\langle \hat{P}_{\mathrm{ans}}\rangle$ for 256 samples with $N=50$.
The inset shows the sample average of $-\log_2\langle\hat{P}_{\mathrm{ans}}\rangle /N$, i.e. 
$\overline{S_{\mathrm{ans}}}/N$ for different sizes. }
\end{figure}
Since the answer fidelity is $0$ for $\lambda=1$ in the thermodynamic
limit and it takes $1$ for $\lambda=0$, it could be a natural candidate for an order parameter.
Indeed, as we can see from Fig.~\ref{fig:Fans}, 
the answer fidelity jumps from almost 0 to a finite value 
in samples exhibiting first order transition.
Also, samples without first order transitions also start to have
non-negligible values from a certain 
value of $\lambda$.
It is tempting to think that the phase transition captured by $\chi_{\mathrm{F}}$
is a transition from a $F_{\mathrm{ans}}=0$ phase to a $F_{\mathrm{ans}}>0$ phase,
but is incorrect.
This is because $F_{\mathrm{ans}}$ seems to converge to 0 for all $\lambda>0$.
In fact,
If we plot $S_{\mathrm{ans}}:=-\log_2\langle \hat{P}_{\mathrm{ans}}\rangle$ normalized by $N$,
they collapse into a common curve as in the inset of
Fig.~\ref{fig:Fans}. 
This is natural, since when $\lambda$ is small enough 
and if $F_{\mathrm{ans}}$ is larger than the fidelity 
between the ground state and any other basis state of the $z$ direction $| {\bm{\sigma}}\rangle$,
$S_{\mathrm{ans}}$ is actually equivalent to the R{\'e}nyi entropy \cite{RenyiEntropy, QRenyiEntropy}
\begin{equation}
S_n(\lambda):=\frac{1}{1-n}\sum_{\bm{\sigma}}\bigl|\langle \bm{\sigma}| \mathrm{GS}(\lambda)\rangle\bigr|^{2n},
\end{equation}
in the $n\rightarrow \infty$ limit,
where $|\mathrm{GS}(\lambda)\rangle$ is the 
ground state of $\hat{H}(\lambda)$.
By assuming that $S_{\infty}/N$ converges to a finite value, 
and that $S_{\infty} = S_{\mathrm{ans}}$, 
it is easy to see that $F_{\mathrm{ans}}$ goes to 0 for all $\lambda>0$.
Similarly to $q$, although both $F_{\mathrm{ans}}$ or $S_{\mathrm{ans}}$
seems to capture the first order transitions, when we take the sample average of them,
no singularities could be observed.

\section{CONCLUSION AND DISCUSSIONS}
We have demonstrated that the fidelity susceptibility can be useful 
to find quantum phase transitions
which would otherwise have been hard to confirm.
For the MIS with unique solutions, 
we find a divergence in $\overline{\chi_{\mathrm{F}}}$, 
which other conventional quantities fail to capture.
Furthermore, previously known first order phase transitions 
in unique solution ensembles of NP hard problems 
occur in the low-$\lambda$ side of this $\overline{\chi_{\mathrm{F}}}$ divergence.
This implies that this ordered phase could be understood as the 
physical consequence of the computational complexity conjecture,
however 
the connections to existing physical scenarios such as spin glass transitions, 
or Anderson/many-body localization are yet to be confirmed.
Further examination of the divergence of $\chi_{\mathrm{F}}$, 
e.g. the critical exponent of $\chi_{\mathrm{F}}$ and its 
connections to other critical exponents,
should be carried out for comparison of different scenarios.

It should be noted that even if quantities like $q$ do not show singularities, 
it does not necessarily mean that the spin glass picture is wrong.
For example, a continuous replica symmetry breaking picture would be compatible 
with no singularities in $\bar{q}$. 
However, to confirm that scenario, one would have to calculate the histogram
$P(q)$ which is rather tedious.
The fidelity susceptibility $\chi_{\mathrm{F}}$ provides an
easier way to confirm the existence of a subtle quantum phase transition.
Consistencies with known freezing phenomena 
found in the actual quantum annealing machine \cite{DWAVEfreeze}
is also an interesting and important problem.

We emphasize that $\chi_{\mathrm{F}}$ also shows a very sharp peak at the first order transitions
which occur within the low $\lambda$ side of the phase transition.  
This suggests that $\chi_{\mathrm{F}}$ may actually detect all of the transitions which 
is relevant for QA.
Previous results \cite{prxmadesimple} show that $\chi_{\mathrm{F}}$
could be used as indicators for various transitions including spin liquids etc.
The present work shows that it is also true for the case of QA of NP-hard problems 
which is glassy and has quenched disorders.

\acknowledgements{
We thank Y. Nishikawa for useful discussions.
Numerical simulation in this work has mainly been performed by using the
facility of the Supercomputer Center, Institute for Solid State Physics,
the University of Tokyo.  
This research was supported by the Grants-in-Aid for Scientific Research
from the JSPS, Japan (No. 25120010 and 25610102).  
}

\appendix{
\section{The Unique solution ensemble}
\label{App:A}
Instead of using simple Erd{\"o}s-R{\'e}nyi random graphs 
which have multiple solutions for the MIS,
we use random graphs which have unique solutions.
This ensures that it is always the minimum energy gap $\Delta E_{\min}$ that causes the 
failure of QA.
If there are multiple solutions, a small $\Delta E_{\min}$ would not necessarily imply failure,
since it can still end in the degenerate ground state of $\hat{H}_{\mathrm{P}}$.
To 
generate random-graph ensemble with a unique solution, 
we randomly add 
edges to the Erd{\"o}s-R{\'e}nyi random graph in the following way.
If the original Erd{\"o}s-R{\'e}nyi random graph already has a unique solution,
we can just use it, although the probability of such a graph occurring 
decreases as $N$ increases. 
When the graph has multiple solutions, it means the vertices could be divided into two groups,
namely the backbones and the non-backbones.
The backbones are the vertices which 
are constantly in the independent set or out of it 
through out all the possible solutions.
If a given graph has a unique solution,
all the vertices belong to the 
backbone by definition.
After checking which vertices the backbones are, 
we randomly assign one of the possible solutions, at random.
If there are more than two non-backbones which are inside of the assigned solution,
we add an edge 
between those two vertices.
This makes the chosen solution no longer valid,
while making sure that there are still solutions of the same size.
We continue this process until there are no more pairs of non-backbones inside a particular maximum independent set solution.
If there still remains a non-backbone vertex with no pair,
we randomly choose one backbone vertex within the solution and add an
edge 
with that.
This procedure always decreases the degeneracy.
When the degeneracy is totally removed and the solution is unique, the procedure ends successfully.
If feasible solutions vanish during this procedure,
we discard the 
graph and start all over again. 
We call this stochastically generated ensemble of graphs, the unique solution ensemble in this paper.

\section{The Dynamic Programming Leaf Removal Algorithm}
\begin{figure}[t]
\includegraphics[width=9cm]{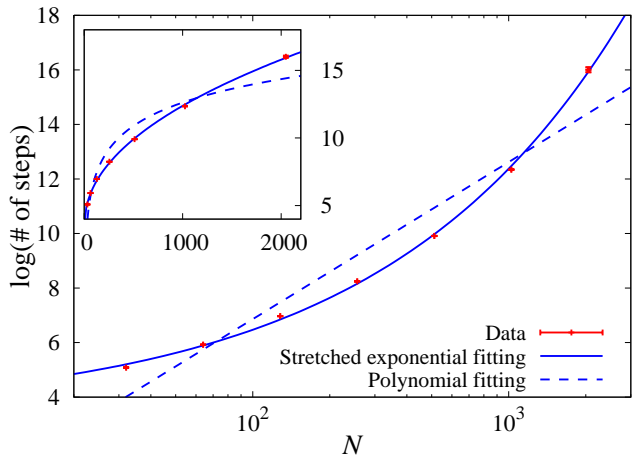}
\caption{\label{fig:DPLRbefore} The logarithm of median time steps
 needed to find the MIS for Erd{\"o}s-R{\'e}nyi random graphs using the
 DPLR algorithm is shown as a function of the system size $N$. The inset
 shows the same plot with non-logarithmic scale for the 
 horizontal axis.
All lines are obtained by least square fittings, either assuming $y=aN^b+c$ (stretched exponential) or 
 $y=c+d\log(N)$ (polynomial).
The stretched exponent $b$ is estimated as $b=0.465(25)$. 
 }
\end{figure}
\begin{figure}[t]
\includegraphics[width=9cm]{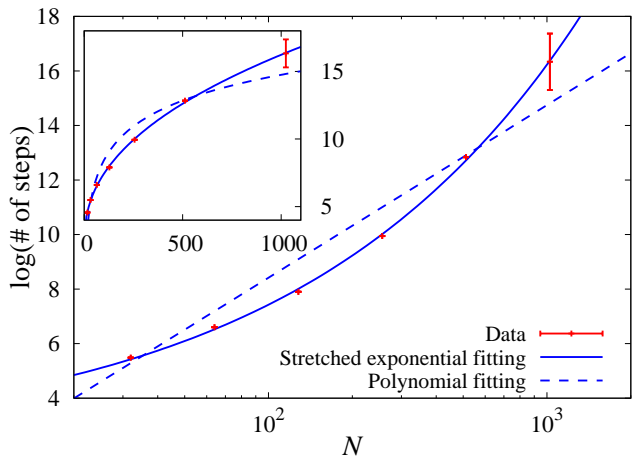}
\caption{\label{fig:DPLRafter} The logarithm of median time steps needed to find the MIS for randomly generated unique solution graphs, using the DPLR algorithm. The inset shows the same plot with non-logarithmic scale for the horizontal axis.
The fittings are performed as the same as in Fig. {\ref{fig:DPLRbefore}},
 yielding that the stretched exponent $b=0.433(19)$. 
}
\end{figure}
We should confirm that the process of making the solution unique does not 
make the problem easier, since we want to know the physical picture of {\it hard} problems. 
A specific algorithm which we call 
the Dynamic Programming Leaf Removal algorithm (DPLR)
was used for confirming this point.
DPLR is a combination of 
leaf removal(LR)\cite{LR}, a standard algorithm for MIS,
and dynamic programming, a well known algorithmic technique.

The LR algorithm is described in Table \ref{Alg: LR}.
We set $k=0$ and $S_v\equiv 0$ for the input. 
They will serve as intermediate 
memories during the recursion of DPLR. 
$k$ represents the largest independent set found so far, 
and $S_v$ represents the configuration corresponding to that.
The LR algorithm only runs until it hits a ``core" where there are no longer vertices with
degree less than 2.

\begin{table}
\begin{algorithm}[H]
  \caption{LR}	
\begin{algorithmic}[ 1]
	\State Input $(G=(V,E), \{S_v\}, k)$
	\While {there exist a vertex $v$ with $S_v=0$ and degree$<2$}
		\State $k++$
		\If {degree of $v=0$}
			\State set $S_v=1$
			\State remove $v$ from the graph $G$
		\EndIf
		\If {degree of $v=1$}
			\State set $S_v=1$
			\State for the vertex $w$ adjacent to $v$, set $S_w=-1$
			\State remove $v$ and $w$ from graph $G$
		\EndIf
	\EndWhile
\State return $(G,\{S_v\}, k)$
 \end{algorithmic}
\end{algorithm}
\caption{\label{Alg: LR} Pseudo code of the LR algorithm.}
\end{table}

We modify the LR algorithm by making leaves, which are vertices with degree
below 2,  with setting a random vertex 
to be either in the independent set or not 
and compare the two results recursively.
The algorithm is presented in Table \ref{Alg: DPLR}.
Again, we set $k=0$ and $S_v\equiv 0$ for the original input. 
$k$ in the final output will represent the size of the maximum independent set,
and $S_v$ in the final output will represent the corresponding configuration.

\begin{table}[t]
\begin{algorithm}[H]
  \caption{DPLR}
   \begin{algorithmic}[ 1]
   \State Input $(G=(V,E), \{S_v\}, k)$
   \State $(G,\{S_v\},k)\leftarrow \mathrm{LR}(G, \{S_v\}, k)$
   \If {$V\neq\varnothing$}
	\State choose one vertex from the remaining graph $v\in V$
	\State $(G^{\prime},\{S^{\prime}_v\},k^{\prime})\leftarrow \mathrm{DPLR}(\mathrm{INC}(G,v), \{S_v\}, k)$
	\State $(G^{\prime\prime},\{S^{\prime\prime}_v\},k^{\prime\prime})\leftarrow \mathrm{DPLR}(\mathrm{EXC}(G,v), \{S_v\}, k)$
		\If{$k^{\prime}+1\geq k^{\prime\prime}$}
			\State\Return $(G^{\prime},\{S^{\prime}_v\},k^{\prime}+1)$
		\Else
			\State\Return  $(G^{\prime\prime},\{S^{\prime\prime}_v\},k^{\prime\prime})$
		\EndIf
	\Else
	\State\Return $(G,\{S_v\},k)$
		\EndIf
\Function{Inc}{$G=(V,E),v$}
\State remove $v$ and adjacent vertices $w\in\partial v$ from graph $G$
\State\Return $G$
\EndFunction
\Function{Exc}{$G=(V,E),v$}
\State remove $v$ from graph $G$
\State\Return $G$
\EndFunction
   \end{algorithmic}
\end{algorithm}
\caption{\label{Alg: DPLR} Pseudo code of the DPLR algorithm.}
\end{table}

It should be noted that this DPLR algorithm is the 
analogue of the well-known DPLL algorithm\cite{DPLL} for the satisfiability problem.
As it is known that the running time of DPLL scales polynomially in the easy parameter region of SAT\cite{SATphasetransition},
DPLR should be a fairly good algorithm to see the hardness of the MIS problem.

We can compare the simple Erd{\"o}s-R{\'e}nyi random graphs and the unique solution ensemble
in terms of computation time using DPLR, from Fig. \ref{fig:DPLRbefore} and Fig. \ref{fig:DPLRafter}.
Since the distribution of time steps has exponentially long tails,
it is convenient to focus on the logarithm of the time steps needed.
We plot the logarithm of the median time steps needed for finding and confirming a solution for the MIS problem.
We used over 1000 samples for each sizes to estimate the median value,
and the error bars are drawn by the bootstrap method. 
Calculating by the logarithm of the time step allows us to have small error bars, 
which otherwise would require an exponential amount of data to have constant size error bars.
Importantly, the median time step shows stretched exponential scaling for both ensembles.
We can see that a polynomial scaling, shown by the dotted lines in
Fig.~\ref{fig:DPLRbefore} and Fig.~\ref{fig:DPLRafter},  does not fit, and a stretched exponential scaling does.
The fact that the median computation time grows stretched exponentially implies that
at least half of the samples will be hard instances.
The exponents are $N^{0.465\pm 0.025}$ and $N^{0.433\pm 0.019}$ for 
Erd{\"o}s-R{\'e}nyi random graphs and the unique solution ensemble respectively.
The fact that both ensembles scale stretched exponentially and not polynomially implies that
the MIS problem is hard for both ensembles and we have not 
changed the hardness of the problem drastically by making the solution
unique described in Appendix~\ref{App:A}.

\nocite{*}

\bibliography{ver2}

\end{document}